\documentclass[prd,twocolumn,superscriptaddress]{revtex4}
\usepackage{amsmath}
\usepackage{adjustbox}
\usepackage{gensymb}
\usepackage{mathrsfs}
\usepackage{amsfonts}
\usepackage{graphicx}
\usepackage{amssymb}
\usepackage{cancel}
\usepackage{color}
\usepackage{amscd}
\usepackage{bm}
\usepackage{adjustbox}
\usepackage{hyperref}
\newcommand{\bs}[1]{\boldsymbol{#1}}
\newcommand{\td}{\tilde{\delta}}
\newcommand{\mc}[1]{\mathcal{#1}}
\makeatletter
\renewcommand{\maketag@@@}[1]{\hbox{\m@th\normalsize\normalfont#1}}%
\makeatletter
\begin{document}

\title{Explicit and covariant formula for thermodynamic volume in extended black hole thermodynamics}
\author{Yong Xiao}
\email{xiaoyong@hbu.edu.cn}
\affiliation{Key Laboratory of High-precision Computation and Application of Quantum Field Theory of Hebei Province,
College of Physical Science and Technology, Hebei University, Baoding 071002, China}
\affiliation{Hebei Research Center of the Basic Discipline for Computational Physics, Baoding, 071002, China}
\affiliation{Higgs Centre for Theoretical Physics, School of Mathematics, University of Edinburgh, Edinburgh, EH9 3FD, United Kingdom}

\author{Yu-Xiao Liu}
\email{liuyx@lzu.edu.cn}
\affiliation{Key Laboratory for Quantum Theory and Applications of the Ministry of Education, Lanzhou Center for Theoretical Physics, Lanzhou University, Lanzhou, Gansu 730000, China}

\author{Yu Tian}
\email{ytian@ucas.ac.cn}
\affiliation{School of Physical Sciences, University of Chinese Academy of Sciences, Beijing 100049, China}
\affiliation{Institute of Theoretical Physics, Chinese Academy of Sciences, Beijing 100190, China}

\author{Hongbao Zhang}
\email{hongbaozhang@bnu.edu.cn}
\affiliation{School of Physics and Astronomy, Beijing Normal University, Beijing 100875, China}
\affiliation{Key Laboratory of Multiscale Spin Physics, Ministry of Education, Beijing Normal University, Beijing 100875, China}

\begin{abstract} In extended black hole thermodynamics, the cosmological constant and other couplings are treated as thermodynamic variables, yielding the first law
$\tilde{\delta}M = T\tilde{\delta}S+\Omega\tilde{\delta}J
+\mathcal{V} \tilde{\delta}P+\cdots$, where $P\equiv -\frac{\Lambda}{8\pi}$. A long-standing conceptual gap in this framework is that, unlike $M$, $T$, $S$, $\Omega$, and $J$, the thermodynamic volume $\mathcal{V} $ lacks a first-principles definition and can only be deduced from other thermodynamic quantities. This deficiency indicates that the underlying origin of $\mathcal{V} $ remains poorly understood. In this paper, we resolve this issue and provide an explicit, covariant formula for $\mathcal{V} $.  
We demonstrate that $\mathcal{V} $ (and the conjugate quantities of other couplings) universally decomposes into two contributions: one arising from the explicit coupling dependence of the Lagrangian, and the other from the response of the fundamental dynamical fields.
This clarifies the physical meaning of the thermodynamic volume and places
it on the same footing as other intrinsic thermodynamic quantities.\end{abstract}

\maketitle



\section{Introduction}

Black hole (BH) thermodynamics has long been a fertile source of new ideas and phenomena since its inception. In asymptotically anti-de Sitter (AdS) spacetimes, the Smarr relation plays a central role in linking different thermodynamic quantities. A particularly striking feature is that the pairs $(\Lambda,\mathcal{V} )$ and $(\alpha_m,\mathcal{V}_m)$, associated with the cosmological constant $\Lambda$ and higher-derivative couplings $\alpha_m$, naturally emerge in the Smarr formula. This observation suggests that $\Lambda$ and $\alpha_m$ should also be treated as thermodynamic variables, leading to the formulation of the extended first law of BH thermodynamics \cite{Kastor:2009wy,Dolan:2010ha,Kubiznak:2012wp,Cvetic:2010jb,Kastor:2010gq,Sinamuli:2017rhp,Dutta:2022wbh,Xiao:2023lap}. At present, extended BH thermodynamics has evolved into a rapidly developing field, with rich connotations and applications across diverse research directions \cite{Wu:2016auq,Frassino:2022zaz,Hajian:2023bhq,Hajian:2021hje,Kubiznak:2016qmn,Mann:2025xrb,Wei:2015iwa,Wei:2019uqg,Bhattacharya:2017hfj,Cong:2021fnf,Ahmed:2023snm,Caceres:2016xjz,Couch:2016exn,AlBalushi:2020rqe,Jacobson:2018ahi,Wei:2026upy,Wei:2023mxw}.

Consider a diffeomorphism-invariant Lagrangian
\begin{equation}
    \mathbf{L}=L\,\boldsymbol{\epsilon}=\Big[\frac{1}{16\pi}(R-2\Lambda)+\sum_m \alpha_m L_m \Big]\boldsymbol{\epsilon},
\end{equation}
where $\boldsymbol{\epsilon}$ denotes the spacetime volume form, $L_m$ represents possible higher-derivative curvature invariants, and $\alpha_m$ are their associated couplings. For a stationary BH solution, the extended first law takes the form
\begin{equation}
    \tilde{\delta}M
    =T\tilde{\delta}S
    +\Omega\tilde{\delta}J
    +\mathcal{V} \tilde{\delta} \Big(-\frac{\Lambda}{8\pi}\Big)
    +\sum_m \mathcal{V}_m\,\tilde{\delta}\alpha_m .
    \label{intro1st}
\end{equation}
Since $-\Lambda/(8\pi)$ is often interpreted as the thermodynamic pressure $P$, the quantity $\mathcal{V} $ is referred to as the thermodynamic volume. Likewise, each $\mathcal{V}_m$ serves as a ``generalized thermodynamic volume" conjugate to the coupling $\alpha_m$.

A key feature of BH thermodynamics is that all \emph{thermodynamic} quantities $M$, $T$, $S$, $\Omega$, and $J$ possess well-defined \emph{geometric} expressions and can be calculated independently. These geometric expressions further reveal that they correspond to intrinsic properties of the system \cite{Bekenstein:1973ur}. Specifically, $T$ is related to surface gravity, while $M$, $J$ and $S$ are associated with spacetime symmetries. However, in Eq.~\eqref{intro1st}, a critical and unsatisfactory fact is that the thermodynamic volumes $\mathcal{V} $ and $\mathcal{V}_m$ are defined solely as derived quantities:
\begin{align}
   \mathcal{V} =\bigg(\frac{\partial M}{\partial P}\bigg)_{S,J,\cdots} \  \text{and} \quad\mathcal{V}_m=\bigg(\frac{\partial M}{\partial \alpha_m}\bigg)_{S,J,\cdots},
\end{align}
and lack a geometric, intrinsic expression. To gain a more complete understanding of extended BH thermodynamics, we therefore address the following question: what are the geometric formulas for $\mathcal{V} $ and $\mathcal{V}_m$, and what intrinsic properties of the system do they reflect?

This has been a persistent concern in the field of extended BH thermodynamics. In the seminal work \cite{Kastor:2009wy}, a geometric expression for $\mathcal{V} $ conjugate to $-\frac{\Lambda}{8\pi}$ was already explored and derived for the simplest Schwarzschild–AdS BH (the concrete expression is given as Eq.~(B1) in Appendix B with further discussion). However, it fails to yield the correct result for more general cases, including the Kerr–AdS BH \cite{Cvetic:2010jb}, the Ba\~nados--Teitelboim--Zanelli (BTZ) BH \cite{Dolan:2010ha,Hajian:2021hje,BTZ:1992}, and other typical BHs in higher-derivative gravity theories \cite{Kastor:2010gq,Sinamuli:2017rhp}. Recently, using the extended Iyer–Wald formalism, Ref.~\cite{Xiao:2023lap} established that Eq.~\eqref{b1} generally accounts for only a part of $\mathcal{V} $ and demonstrated the existence of an additional term. Meanwhile, Ref.~\cite{Hajian:2023bhq} provided an alternative interpretation of this additional term as a pure gauge term (subject to gauge fixing) of an auxiliary gauge field in an equivalent Lagrangian formulation. Nevertheless, the calculations in Refs.~\cite{Xiao:2023lap,Hajian:2023bhq} still rely on integrability arguments involving other quantities such as $\tilde\delta M$. Consequently, deriving an explicit, independent geometric formula for $\mathcal{V} $ and $\mathcal{V}_m$ and clarifying its fundamental physical origin remains an open problem. 

In fact, identifying a suitable theoretical framework is crucial for resolving this issue. Most recently, substantial progress has been made in understanding conventional BH thermodynamics; see the series of works by some of the present authors, Zhang and Xiao \cite{Guo:2025ohn,Guo:2025muo,Chen:2025ary,Xiao:2025icr}.  In this paper, we show that, once the newly developed methodology is applied to extended BH thermodynamics, all the geometric terms in the Iyer--Wald charge identity are almost automatically separated. Interestingly, these terms correspond one-to-one to thermodynamic quantities (including $\mathcal{V}_i$ that we concern), as clearly visualized in Eq.~\eqref{sec2firstlawE}. Our finding uncovers a universal decomposition of $\mathcal{V}_i$ of the form:
\begin{align}
    \mathcal{V}_i = \mathcal{V}_i^{(1)}+\mathcal{V}_i^{(2)}. \label{sec2v1v2}
\end{align}
 where $\mathcal{V}_i^{(1)}$ and $\mathcal{V}_i^{(2)}$ are given in Eqs.~\eqref{resv1} and \eqref{resv2} and have a clear physical interpretation. For the sake of unified notation, we have taken $i=(0,m)$, with $\mathcal{V}_0=\mathcal{V} $ and $\alpha_0=-\frac{\Lambda}{8\pi}$.

This paper is organized as follows. The derivation of the explicit covariant formula for the thermodynamic volume $\mathcal{V}_i$ is presented in Sec.~\ref{sec2}. To verify its validity, we apply the formula to two concrete examples in Sec.~\ref{sec3} and Sec.~\ref{sec4}. Our concluding remarks are given in Sec.~\ref{sec5}. Throughout this work, we use the background subtraction method to regularize the divergences of AdS quantities. The alternative regularization scheme, holographic renormalization, is discussed in Appendix A. Note that our work represents a distinct technical pathway compared to existing attempts in this direction, so a detailed comparison is provided in Appendix B.

\section{Derivation of the formulas for the thermodynamic volume}\label{sec2}

We will provide a concise review of the Iyer--Wald formalism \cite{Wald:1993nt,Iyer:1994ys} and its extended version \cite{Caceres:2016xjz,Couch:2016exn,Xiao:2023lap} to ensure self-containedness. We use $\delta$ and $\tilde{\delta}$ to denote variations in the conventional and extended BH thermodynamics, respectively. Then we incorporate recent progress \cite{Guo:2025ohn,Guo:2025muo,Chen:2025ary,Xiao:2025icr} and obtain explicit formulas for $\mathcal{V}_i$.

The variation of the Lagrangian $\mathbf{L}$ with respect to the dynamical fields $\phi \equiv \{g_{\mu\nu},\psi\}$ (where $\psi$ represents possible matter fields) takes the form
\begin{equation}
    \delta\mathbf{L} = \mathbf{E}^{\phi}\,\delta\phi + d\boldsymbol{\Theta}[\delta\phi],
    \label{gelvariform}
\end{equation}
where $\mathbf{E}^{\phi} = 0$ is the equation of motion, and $d\boldsymbol{\Theta}$ denotes a total derivative term. For a fixed vector field $\xi$, the Noether current is defined as
$\mathbf{J}_\xi \equiv \boldsymbol{\Theta}[\mathcal{L}_\xi\phi] - \xi\cdot\mathbf{L}$, where $\mathcal{L}_\xi$ denotes the Lie derivative along $\xi$. Since $d\mathbf{J}_\xi = 0$ on-shell, we can construct the Noether charge $\mathbf{Q}_\xi$ such that $\mathbf{J}_\xi = d\mathbf{Q}_\xi$. When $\xi$ is a Killing vector, the Iyer--Wald formalism establishes a pivotal identity:
\begin{align}
    d \boldsymbol{k}_\xi = 0,
    \label{ff1}
\end{align}
where $\boldsymbol{k}_\xi$ is the Iyer--Wald surface charge density defined by
\begin{equation}
    \boldsymbol{k}_\xi
    \equiv
    \delta\mathbf{Q}_{\xi}
    -
    \xi\cdot\boldsymbol{\Theta}[\delta\phi]. \label{sec2wdef}
\end{equation}
Eq.~\eqref{ff1} is proven under the conditions $\mathbf{E}^\phi=0$ and $\delta\mathbf{E}^\phi=0$, meaning the variation in Eq.~\eqref{ff1} should be interpreted as a variation within the solution space, typically parameterized by quantities such as the mass parameter $m$, rotation parameter $a$, or conserved charge $q$.

For a given stationary BH, the horizon Killing vector is written as $\xi_H=\xi_t+\Omega_H \xi_\phi$, where $\xi_t\equiv \partial_t$, $\xi_\phi \equiv \partial_\phi$, and $\Omega_H$ is the angular velocity at the horizon. Here, we work in four-dimensional spacetime with coordinates $(t,r,\theta,\phi)$; generalization to other spacetime dimensions is straightforward. Integrating the identity $d\boldsymbol{k}_{\xi_H}=0$ over a spatial hypersurface $\Sigma$ that extends from the bifurcation surface $S_{r_h}$ (with $r_h$ the horizon radius) to the surface $S_\infty$ (the limit of $S_{r_c}$ as $r_c\rightarrow\infty$, representing a radial cutoff), the generalized Stokes' theorem yields
\begin{equation}
   \int_{S_\infty}\boldsymbol{k}_{\xi_H}- \int_{S_{r_h}}\boldsymbol{k}_{\xi_H} = 0.
    \label{id1}
\end{equation}
Then we evaluate $\boldsymbol{\Theta}$ on the timelike boundary $\partial \mathcal{M}$ at infinity, where $S_\infty = \Sigma \cap \partial \mathcal{M}$. The induced volume form $\hat{\boldsymbol{\epsilon}}$ on $\partial\mathcal{M}$ is defined by $\boldsymbol{\epsilon} = \boldsymbol{n}\wedge \hat{\boldsymbol{\epsilon}}$, with $\boldsymbol{n}\propto \nabla r$ being the outward unit normal vector to $\partial \mathcal{M}$. Following standard procedures of \cite{Padmanabhan:2014lwa,Parattu:2015gga,Jiang:2018sqj}, Refs.~\cite{Guo:2025ohn,Guo:2025muo,Chen:2025ary,Xiao:2025icr} suggest decomposing  
\begin{equation}
    \boldsymbol{\Theta}[\delta\phi]\big|_{\partial\mathcal{M}} = -\delta\boldsymbol{B} + d\boldsymbol{C}[\delta\phi] + \boldsymbol{F}[\delta\phi]. \label{sec2theta1}
\end{equation}
Here, $\boldsymbol{B}$ is the boundary term associated with the Lagrangian $\boldsymbol{L}$; for example, it is the Gibbons-Hawking-York term in Einstein gravity. The $d\boldsymbol{C}$ term can be safely omitted for simplicity \cite{Parattu:2015gga,Guo:2024oey,footnote1}. The remaining term $\boldsymbol{F}[\delta\phi]$ is a familiar quantity in action variations. Combining Eqs.~\eqref{gelvariform} and \eqref{sec2theta1}, the variation of the complete action including boundary terms yields 
\begin{align}
    \delta \left( \int_{\mathcal{M}} \mathbf{L} + \int_{\partial\mathcal{M}} \mathbf{B} \right) \sim \int_{\mathcal{M}} \mathbf{E}^{\phi} \delta\phi + \int_{\partial\mathcal{M}} \boldsymbol{F}[\delta\phi],\label{vlb}
\end{align}
where $\boldsymbol{F}[\delta\phi]$ has a typical form such as $T_{ij}\delta h^{ij} + \cdots$. For instance, in Einstein gravity, we have $\boldsymbol{F}[\delta \phi] = \frac{1}{16 \pi}(K_{\mu \nu} - K h_{\mu\nu}){\delta} h^{\mu\nu}\hat{\boldsymbol{\epsilon}}$, where $(K_{\mu \nu} - K h_{\mu\nu})$ corresponds to the Brown--York tensor and ${\delta} h^{\mu\nu}$ denotes the variation of the induced metric on the boundary $\partial \mathcal{M}$. In general, $\boldsymbol{F}[\delta \phi]$ involves the dynamical fields that must be fixed when deriving the equations of motion $\mathbf{E}^{\phi}=0$ (though these fields should not be fixed when computing conserved charges). As explained in Ref.~\cite{Parattu:2015gga}, the fixed fields represent the true fundamental degrees of freedom of the system.

In the Iyer--Wald formalism, $\int_{S_{r_h}}\boldsymbol{k}_{\xi_H}$ is identified as $T\,\delta S$, where $T$ is the Hawking temperature and $S$ is the BH entropy \cite{Wald:1993nt,Iyer:1994ys}. Substituting it and Eqs.~\eqref{sec2wdef} and \eqref{sec2theta1} into Eq.~\eqref{id1}, we obtain
\begin{equation}
\int_{S_\infty}^{(\mathrm{BH})} \left( \delta\boldsymbol{Q}_{\xi_H} + \xi_H\cdot\delta\boldsymbol{B}  - \xi_H\cdot\boldsymbol{F}[\delta\phi] \right) - T\delta S = 0.
    \label{iwadsbh}
\end{equation}
The integral of the terms in the parentheses is individually divergent, and we will adopt a background subtraction procedure to regularize them. This prescription originates from the foundational work \cite{hawkingpage}, based on the principle that a well-defined physical quantity should measure the BH value relative to the AdS vacuum, rather than a bare divergent quantity. In the standard background subtraction scheme \cite{hawkingpage,Gubser:1998nz,Gibbons:2004ai,Dutta:2006vs}, the reference AdS spacetime is redshifted via $t\rightarrow \lambda(r_c)t$, motivated by ensuring that the BH and the background share identical asymptotic geometries at infinity. It is important to note that, throughout this paper, we adopt the \emph{redshifted} background subtraction, which is distinct from the \emph{unredshifted} subtraction discussed in Appendix B.

For the horizonless background spacetime, integrating Eq.~\eqref{ff1} gives $\int_{S_\infty}\boldsymbol{k}_{\xi_H} = 0$, leading to
\begin{equation}
    \int_{S_\infty}^{(\mathrm{bg})} \left( \delta\boldsymbol{Q}_{\xi_H} + \xi_H\cdot\delta\boldsymbol{B}  - \xi_H\cdot\boldsymbol{F}[\delta\phi] \right) = 0.
    \label{iwads0}
\end{equation}
Subtracting Eq.~\eqref{iwads0} from Eq.~\eqref{iwadsbh}, we arrive at 
\begin{align}
\begin{split}
  \int_{S_\infty}^{(\mathrm{Reg.})} & \Big[\overbrace{\left( \delta\boldsymbol{Q}_{\xi_t}  + \Omega_\infty \delta \boldsymbol{Q}_{\xi_\phi} + \xi_t\cdot\delta\boldsymbol{B} \right)}^{\delta M} + \overbrace{\Omega \,\delta \boldsymbol{Q}_{\xi_\phi}}^{-\Omega \,\delta J}\Big] \\
  & - T\,\delta S  - \int_{S_\infty}^{(\mathrm{Reg.})} \overbrace{\xi_t\cdot\boldsymbol{F}[\delta\phi]}^{0} = 0,
    \label{sec2firstlaw}
    \end{split}
\end{align}
where we use $\int_{S_\infty}^{(\mathrm{Reg.})}$ to abbreviate $\int_{S_\infty}^{(\mathrm{BH})} - \int_{S_\infty}^{(\mathrm{bg})}$. As shown in Refs.~\cite{Guo:2025ohn,Guo:2025muo,Chen:2025ary,Xiao:2025icr}, the redshifted AdS background ensures the cancellation of the $\boldsymbol{F}$ terms. Some technical details: when an AdS BH has a nonzero angular velocity $\Omega_\infty$ at infinity, the Killing vector associated with $\delta M$ takes the form $\partial_t + \Omega_\infty \partial_\phi$ \cite{Xiao:2023lap,Gibbons:2004ai}. Thus, at infinity, we decompose $\xi_H$ as $\xi_H = (\xi_t + \Omega_\infty \xi_\phi) + \Omega \xi_\phi$, where $\Omega = \Omega_H - \Omega_\infty$. Additionally, since $\xi_\phi$ is tangent to $S_\infty$, the terms involving $\int_{S_\infty} \xi_H \cdot \hat{\boldsymbol{\epsilon}}$ can be rewritten as $\int_{S_\infty} \xi_t \cdot \hat{\boldsymbol{\epsilon}}$.

By comparing the first law $\delta M - \Omega \delta J - T\delta S = 0$ with the geometric relation \eqref{sec2firstlaw}, we readily recognize the expressions for mass $M$ and angular momentum $J$:
\begin{align}
   & M = \int_{S_\infty}^{(\mathrm{Reg.})} \left( \boldsymbol{Q}_{\xi_t} + \Omega_\infty \boldsymbol{Q}_{\xi_\phi} + \xi_t \cdot \boldsymbol{B} \right), \label{masscon} \\
  & J = -\int_{S_\infty}^{(\mathrm{BH})} \boldsymbol{Q}_{\xi_\phi}, \label{angularcon}
\end{align}
and also the standard Wald entropy formula as given in \cite{Wald:1993nt,Iyer:1994ys}. The mass formula above can be further rewritten in the ADM or Brown-York form \cite{Iyer:1994ys,Harlow:2019yfa,Guo:2024oey}. We omit the $^{(\mathrm{Reg.})}$ symbol in Eq.~\eqref{angularcon} because the background contribution to $J$ vanishes.

A caveat applies to rotating AdS BHs: in extracting Eq.~\eqref{masscon} from Eq.~\eqref{sec2firstlaw}, we implicitly assumed $\delta \Omega_\infty = 0$. This means that both variations and background subtraction must be performed within a solution space that shares the same $\Omega_\infty$. This requirement is easily satisfied. For example, consider a Kerr-AdS BH where $\Omega_\infty = -\frac{a}{l^2}$. If we perform variations $m \rightarrow m + \delta m$ and $a \rightarrow a + \delta a$, the corresponding $\Omega_\infty$ becomes $\Omega_\infty' = -\frac{a + \delta a}{l^2}$. To restore $\Omega_\infty$ to its original value, we can introduce an additional coordinate transformation: $t \rightarrow t$ and $\phi \rightarrow \phi - \frac{\delta a}{l^2} t$. Notably, this requirement of fixing $\Omega_\infty$ is consistent with the well-known fact that when calculating the Euclidean action for a Kerr-AdS BH, the background spacetime should be chosen with $m=0$, rather than $m=0$ and $a=0$.

Next, we extend the formalism by introducing the operator $\tilde{\delta}$, which allows variations of the couplings $\alpha_i$. For the Lagrangian $\mathbf{L} = ( \frac{1}{16\pi}R + \sum_i \alpha_i L_i ) \bs{\epsilon}$, the extended variation of $\mathbf{L}$ with respect to the dynamical fields $\phi \equiv \{g_{\mu\nu},\psi\}$ is
\begin{align}
\begin{split}
     \tilde{\delta} \mathbf{L} &= \frac{\partial \mathbf{L}}{\partial \phi}\tilde{\delta} \phi + \sum_i \frac{\partial \mathbf{L}}{\partial \alpha_i}\tilde{\delta} \alpha_i \\
     &= \mathbf{E}^{\phi} \tilde{\delta} \phi + d \boldsymbol{\Theta}[\tilde{\delta} \phi] + \sum_i L_i \boldsymbol{\epsilon} \, \tilde{\delta} \alpha_i. \label{extendedDL}
\end{split}
\end{align}
The last term arises from the explicit dependence of $\mathbf{L}$ on $\alpha_i$. As a consequence, in the extended formalism \cite{Xiao:2023lap}, Eq.~\eqref{ff1} generalizes to 
\begin{align}
    d\boldsymbol{k}_\xi = - \sum_i \xi \cdot ( L_i \boldsymbol{\epsilon} \, \tilde{\delta}\alpha_i).
\end{align}
Applying the horizon Killing vector $\xi_H$, we integrate this identity over the hypersurface $\Sigma$ and follow the same reasoning as before, with the decomposition of $\boldsymbol{\Theta}[\tilde{\delta} \phi]$ included. The extended version of Eq.~\eqref{sec2firstlaw} then yields
\begin{align}
\begin{split}
   &\int_{S_\infty}^{(\mathrm{Reg.})} [\overbrace{\tilde{\delta}\boldsymbol{Q}_{\xi_t} + \Omega_\infty \tilde{\delta} \boldsymbol{Q}_{\xi_\phi} + \xi_t \cdot \tilde{\delta}\boldsymbol{B}}^{\tilde{\delta} M} + \overbrace{\Omega \, \tilde{\delta} \boldsymbol{Q}_{\xi_\phi}}^{-\Omega \, \tilde{\delta} J} ] - T\,\tilde{\delta} S  \\
  & \quad - \int_{S_\infty}^{(\mathrm{Reg.})} \overbrace{\xi_t \cdot \boldsymbol{F}[\tilde{\delta}\phi]}^{\mathcal{V}_i^{(2)} \tilde{\delta} \alpha_i} = \sum_i \int_{\Sigma}^{(\mathrm{Reg.})} \overbrace{(-L_i) \xi_t \cdot \boldsymbol{\epsilon} \, \tilde{\delta} \alpha_i}^{\mathcal{V}_i^{(1)} \tilde{\delta}\alpha_i}.
    \end{split}\label{sec2firstlawE}
\end{align}
By comparing the extended first law $\tilde{\delta} M = T\tilde{\delta} S+\Omega \tilde{\delta} J +\mathcal{V}_i \tilde{\delta} \alpha_i $ with the geometric relation \eqref{sec2firstlawE}, we still identify the mass and angular momentum as Eqs.~\eqref{masscon} and \eqref{angularcon}. However, we now clearly extract the formula for $\mathcal{V}_i$, which contains two parts: $\mathcal{V}_i = \mathcal{V}_i^{(1)} + \mathcal{V}_i^{(2)}$. The expressions for $\mathcal{V}_i^{(1)}$ and $\mathcal{V}_i^{(2)}$ are recognized as
\begin{align}
  &  \mathcal{V}_i^{(1)} = -\int_{\Sigma}^{(\mathrm{Reg.})} L_i \, \xi_t \cdot \boldsymbol{\epsilon}, \label{resv1} \\
    & \mathcal{V}_i^{(2)} = \int_{S_\infty}^{(\mathrm{Reg.})} F[ \tilde{\delta} \phi /\tilde{\delta} \alpha_i ] \xi_t \cdot \hat{\boldsymbol{\epsilon}}. \label{resv2}
\end{align}
We now analyze the physical origins of $\mathcal{V}_i^{(1)}$ and $\mathcal{V}_i^{(2)}$. First, $\mathcal{V}_i^{(1)}$ clearly arises from the last term in Eq.~\eqref{extendedDL}, reflecting the system's response to $\alpha_i$ variations that stems from the inherent form of the Lagrangian $\boldsymbol{L}$ \cite{footnote2}. Next, for $\mathcal{V}_i^{(2)}$, we clarified around Eq.~\eqref{vlb} that $\boldsymbol{F}[\delta \phi]$ originates from action variations due to changes in the system's fundamental dynamical degrees of freedom. Thus, the term $\boldsymbol{F}[ \tilde{\delta} \phi /\tilde{\delta} \alpha_i ]$ describes the response of these degrees of freedom to coupling variations.

Our formulas \eqref{resv1} and \eqref{resv2} are covariant and can be explicitly calculated: the Lagrangian determines their concrete geometric expression, and the BH metric fixes the final calculation result. The symbol ``$\int^\mathrm{(Reg.)}$" in the mass formula and volume formulas indicates that the physical quantities are extracted via a regularization procedure. A reasonable regularization must reproduce the correct mass of the BH, the only potential ``regularization dependence'' emerges when vacuum energy is involved. Specifically, when switching the regularization scheme from background subtraction \cite{hawkingpage,Gubser:1998nz,Gibbons:2004ai,Dutta:2006vs} to holographic renormalization \cite{Balasubramanian:1999re,Bianchi:2001kw,deHaro:2000vlm}, the latter yields the Casimir energy $E_C$ in odd spacetime dimensions, while the former (which measures only relative quantities) does not. In such cases, the mass given by Eq.~\eqref{masscon}, as well as the two contributions to the volume from Eqs.~\eqref{resv1} and \eqref{resv2}, shift in a correlated manner. Nevertheless, the extended first law itself remains invariant under such scheme changes. The appearance of ``$\int^\mathrm{(Reg.)}$'' on both sides of Eq.~\eqref{sec2firstlawE} actually reflects the self-consistency of our formalism.

\section{Example I: Thermodynamic volume for Kerr-AdS BH} \label{sec3}

Having derived explicit formulas for $\mathcal{V}_i$ through general analysis, we now verify their correctness by applying them to representative examples.

The Lagrangian of Einstein gravity is given by $ \bs{L}=\frac{1}{16\pi}(R-2\Lambda)\bs{\epsilon}$. The Kerr-AdS BH is the rotating solution of this theory. Its metric in Boyer-Lindquist coordinates is \cite{Gibbons:2004ai}
\begin{align}
\begin{split}
  ds^2 =& \!-\! \frac{\Delta_r}{\rho^2}\left(dt\! -\! \frac{a\sin^2\theta}{\Xi}\,d\phi\right)^2 \!+ \!\frac{\rho^2}{\Delta_r}\,dr^2 \!+\! \frac{\rho^2}{\Delta_\theta}\,d\theta^2\\ & + \frac{\Delta_\theta\sin^2\theta}{\rho^2}
  \left(a\,dt - \frac{r^2+a^2}{\Xi}\,d\phi\right)^2, \label{kerrmetric}
  \end{split}
\end{align}
where $ \Delta_r = (r^2+a^2)(1+\frac{r^2}{l^2})-2mr$,
 $\Delta_\theta = 1-\frac{a^2}{l^2}\cos^2\theta$,
$\rho^2 =\ r^2+a^2\cos^2\theta$, and $\Xi = 1-\frac{a^2}{l^2}$. The AdS length $l$ is defined via $\Lambda=-\frac{3}{l^2}$ in $D=4$ dimensions. This metric has a nonzero angular velocity $\Omega_\infty=-a/l^2$ at infinity.

 For Einstein gravity, the decomposition of $\bs{\Theta}$ is familiar \cite{Padmanabhan:2014lwa,Parattu:2015gga}:
\begin{equation}
    \boldsymbol{\Theta}[\td\phi]\big|_{\partial\mc{M} } \!= \!\frac{1}{16\pi}\left[ \!-\! \td(2K\hat{\boldsymbol{\epsilon}}) + (K_{\mu\nu} \!- \!Kh_{\mu\nu})\td h^{\mu\nu}\hat{\boldsymbol{\epsilon}} \right].\label{thEinstein}
\end{equation}
Here, $h_{\mu\nu}\equiv g_{\mu\nu}-n_\mu n_\nu$ is the induced metric on $\partial\mc{M}$, $K_{\mu\nu}=h_{\mu}^{\ \rho}\nabla_{\rho}n_{\nu}$ is the extrinsic curvature tensor, and $K$ is its trace. From Eq.~\eqref{thEinstein}, we can directly identify the $\boldsymbol{F}\equiv F\, \hat{\boldsymbol{\epsilon}}$ term with
\begin{equation}
    F[\tilde{\delta} \phi] = \frac{1}{16\pi} (K_{\mu\nu} - Kh_{\mu\nu}) \tilde{\delta} h^{\mu\nu},\label{einF}
\end{equation}
which depends only on the variation of the induced metric.

Considering the variation of $\Lambda$, the extended first law for Kerr-AdS BHs then takes the form
\begin{align}
    \tilde{\delta} M = T \tilde{\delta} S + \Omega \tilde{\delta} J + \mathcal{V}  \tilde{\delta}\left(-\frac{\Lambda}{8\pi}\right).
\end{align}
Now, all the physical quantities in the thermodynamic relation have well-established, geometry-based expressions. We will focus on verifying our formulas for the thermodynamic volume $\mathcal{V} $.

Using the formula \eqref{resv1} with $i=0$ (where $\alpha_0\equiv-\frac{\Lambda}{8\pi}$, $L_0\equiv 1$), the first part of $\mathcal{V} $ can be easily calculated as
\begin{align}
    \mathcal{V} ^{(1)}\!=\!-\!\int_{\Sigma}^{(\mathrm{Reg.})}\xi_t \cdot \bs{\epsilon}\!=\! \frac{4 \pi  l^2 r_h  \left(a^2 \!+\! r_h ^2\right)}{3 \left(l^2\!-\! a^2\right)}\!+\! \frac{4 \pi  l^4 m}{3 a^2 \!-\! 3 l^2},
\end{align}
where $\xi_t=\partial_t=\{1,0,0,0\}$, and $\bs{\epsilon}$ is the volume form of the space-time. Meanwhile, using the formula \eqref{resv2} and  the expression \eqref{einF} for $\bs{F}$, a direct calculation produces
\begin{align}
    & \mathcal{V} ^{(2)}=\int_{S_\infty}^{(\mathrm{Reg.})} F\big[\td h^{\mu\nu}/\td \alpha_0 \big] \xi_t \cdot\hat{\bs{\epsilon}}=\frac{4 \pi  l^6 m}{3 \left(a^2-l^2\right)^2}.
\end{align}
The background metric is constructed in two steps: setting $m=0$ in Eq.~\eqref{kerrmetric} (yielding pure AdS spacetime in rotating coordinates) and introducing a redshift factor via $t\rightarrow \sqrt{ \frac{\Delta_r(r_c)}{\Delta_r(r_c)|_{m=0}} }\, t$. Then, the background subtraction procedure is straightforward: first evaluate the relevant integrals from the inner boundary to a cutoff $r_c$ for both the Kerr--AdS geometry and the background; then subtract these two results, so that the divergence at $r_c$ cancels out, allowing us to safely take $r_c\to\infty$.

Adding $\mathcal{V} ^{(1)}$ and $\mathcal{V} ^{(2)}$ together, we obtain
\begin{align}
    \mathcal{V} =\mathcal{V} ^{(1)}+\mathcal{V} ^{(2)}=\frac{4 \pi   r_h  \left(a^2 + r_h ^2\right)}{3 \left(1 - \frac{a^2}{l^2}\right)}+\frac{4 \pi  a^2 m}{3 \left(1 - \frac{a^2}{l^2}\right)^2}.\label{kerrv}
\end{align}
This result coincides exactly with the known expressions \cite{Cvetic:2010jb,Xiao:2023lap}, previously abbreviated as $\mathcal{V} =V+\frac{4\pi}{3}Ma^2$. Furthermore, taking $a\rightarrow 0$, it reduces to the familiar thermodynamic volume for Schwarzschild-AdS BHs: $\mathcal{V} =\frac{4\pi}{3}r_h^3$.

\section{Example II: Thermodynamic volume for rotating BTZ BHs in new massive gravity}  \label{sec4}

We next consider another example: a theory with higher-derivative terms that introduce additional coupling parameters beyond $\Lambda$. This illustrates the generality of our formalism for the extended thermodynamic pairs $(\alpha_m,\mathcal{V}_m)$.

 The Lagrangian of the new massive gravity in $D=3$ dimensions is given by \cite{Bergshoeff:2009hq,Hajian:2023bhq}
\begin{align}
    \mathcal{L}=\frac{1}{16\pi} \left[R-2\Lambda-\alpha\left(\frac{3}{8}R^2-R_{\mu\nu}R^{\mu \nu}\right)\right].
\end{align}
where $\alpha$ denotes the corresponding higher-derivative coupling. The rotating BTZ BH solution reads \cite{BTZ:1992,Clement:2009}
\begin{align}
    ds^2=-\Delta dt^2+\frac{dr^2}{\Delta}+r^2\left(d \varphi-\frac{j}{2r^2} dt\right)^2, \label{3dmetric}
\end{align}
where $\Delta\equiv -m+\frac{r^2}{l^2}+\frac{j^2}{4r^2}$. The relation between $\Lambda$ and the effective AdS length $l$ is modified by $\alpha$, which is $\Lambda=-\frac{1}{l^2}+\frac{\alpha}{4l^4}$. Notably, this spacetime has $\Omega_\infty=0$, avoiding rotational subtleties at infinity.

For $f(R_{\mu\nu\rho\sigma})$ gravity, the $\bs{F}\equiv F\,\hat{\bs{\epsilon}}$ term has a known general form \cite{Jiang:2018sqj,Guo:2025ohn}:
\begin{align}
& F[\td\phi] = 4K_{\alpha\beta}\td \mathcal{P}^{\alpha\beta} + 2\Big[n^{\nu}\nabla^{\mu}P_{\alpha\mu\nu\beta} -\mathcal{P}_{\mu\alpha}K^{\mu}_{\ \beta} \nonumber \\
&\quad - \mathcal{P}_{\mu\nu}K^{\mu\nu}h_{\alpha\beta}  - D^{\mu}(h_{\mu}^{\ \nu} h_\alpha^{\rho}P_{\nu \rho \sigma\beta}n^{\sigma})   \Big] \td  h^{\alpha\beta} ,  \label{3dfterm}
\end{align}
where $P^{\mu\nu\rho\sigma}\equiv \frac{\partial L}{\partial R_{\mu\nu\rho\sigma}}$ (Lagrangian derivative w.r.t. Riemann tensor) and $\mathcal{P}^{\mu\nu}\equiv P^{\alpha\beta\gamma\delta}n_{\beta}n_{\delta}h_{\alpha}^{\ \mu}h_{\gamma}^{\ \nu}$ (projected tensor). For the present case, we obtain
\begin{align}
\begin{split}
  P^{\mu\nu\rho\sigma}  =\frac{1}{16\pi}\bigg( g^{\mu[\rho}g^{\sigma]\nu} (1-\frac{3\alpha R}{4})+ 2 \alpha g^{\mu[\rho}R^{\sigma]\nu}  \bigg).
\end{split}
\end{align}
Substituting this into Eq.~\eqref{3dfterm} determines the $\boldsymbol{F}$ term for our calculations.

The extended first law for this system includes both $\Lambda$ and $\alpha$ as thermodynamic variables:
\begin{align}
    \delta M=T \delta S+ \Omega dJ+\mathcal{V} \delta \left(-\frac{\Lambda}{8\pi} \right)+\mathcal{V}_\alpha \delta \alpha.
\end{align}
We first calculate the thermodynamic volume conjugate to $\alpha_0\equiv -\frac{\Lambda}{8\pi}$, which gives
\begin{align}
  &  \mathcal{V} ^{(1)}=-\int_{\Sigma}^{(\mathrm{Reg.})}\xi_t\cdot \bs{\epsilon}=\pi  r_h ^2-\frac{1}{2} \pi  l^2 m,\\
& \mathcal{V} ^{(2)}=\int_{S_\infty}^{(\mathrm{Reg.})} F\big[\td \phi/{\td \alpha_0} \big] \xi_t \cdot\hat{\bs{\epsilon}}=\frac{\pi  l^2 m \left(2 l^2+\alpha \right)}{2 \left(2 l^2-\alpha \right)}.
\end{align}
In the above subtraction procedure, the background metric is obtained as follows: setting $m=0$ and $j=0$ in Eq.~\eqref{3dmetric} yields the pure AdS metric, which is then redshifted via the transformation $t\rightarrow \sqrt{\frac{\Delta(r_c)}{\Delta(r_c)|_{m=j=0}} }\, t$. Adding the two components gives the following result:
\begin{align}
    \mathcal{V}  = \mathcal{V} ^{(1)} + \mathcal{V} ^{(2)} = \pi r_h^2 + \frac{\pi \alpha l^2 m}{2l^2 - \alpha}.
\end{align}

Next, we calculate the conjugate quantity $\mathcal{V}_\alpha$ with  $L_\alpha=\frac{1}{16\pi} (-\frac{3}{8}R^2+R_{\mu\nu}R^{\mu \nu})$, which yields
\begin{align}
  &   \mathcal{V}_\alpha^{(1)}=-\int_{\Sigma}^{(\mathrm{Reg.})}L_\alpha \,\xi_t\cdot \bs{\epsilon}=\frac{3 \left(l^2 m-2 r_h ^2\right)}{64 l^4},\\
&\mathcal{V}_\alpha^{(2)}=\int_{S_\infty}^{(\mathrm{Reg.})} F\big[\td \phi/{\td \alpha} \big] \xi_t \cdot\hat{\bs{\epsilon}}=\frac{m \left(2 l^2+\alpha \right)}{64 l^2 \left( 2 l^2-\alpha\right)}.
\end{align}
 Adding them together, we obtain the result
\begin{align}
   \mathcal{V}_\alpha=-\frac{3 r_h ^2}{32 l^4} +\frac{m \left(4 l^2-\alpha \right)}{32 l^2 \left(2 l^2-\alpha \right)}.
  \end{align}
Both $\mathcal{V} $ and $\mathcal{V}_\alpha$ match the known results in \cite{Hajian:2023bhq}, further confirming the validity of our formula.

\section{Concluding remarks}  \label{sec5}

In this paper, we resolve an open issue in extended BH thermodynamics: thermodynamic volumes $\mathcal{V}_i$ were previously defined only via thermodynamic relations, lacking an independent geometric expression. This imbalance is conceptually unsatisfactory, as it distinguishes $\mathcal{V}_i$ from other fundamental thermodynamic quantities. 

Two key ingredients from the newly developed framework \cite{Guo:2025ohn,Guo:2025muo,Chen:2025ary,Xiao:2025icr}
allow us to overcome the difficulties in the previous literature and achieve our results.
First, we employ the decomposition
$\boldsymbol{\Theta}|_{\partial \mathcal{M}}= -\delta\boldsymbol{B} + d\boldsymbol{C} + \boldsymbol{F}$,
which enables us to resolve finer structures of $\boldsymbol{\Theta}$ within the Iyer--Wald formalism.
Second, the framework \cite{Guo:2025ohn,Guo:2025muo,Chen:2025ary,Xiao:2025icr}
naturally incorporates the standard regularization scheme in BH thermodynamics,
either redshifted background subtraction or holographic renormalization.
Within this consistent framework, all geometric terms arrange neatly in Eq.~\eqref{sec2firstlawE}, and consequently an explicit and covariant formula for the thermodynamic volume emerges automatically.

The formula shows that $\mathcal{V}_i$ corresponds to an intrinsic property of the system: it encodes two distinct responses to coupling variations, originating from the Lagrangian structure and fundamental dynamical fields. This effort of relating $\mathcal{V}_i$ to a fundamental property is in parallel with identifying BH temperature $T$ with surface gravity, and identifying $M$ and $J$ with conserved charges of spacetime symmetries. Our results therefore provide a complete physical interpretation of each term in the extended first law.

While focused on pure gravity here, our method is flexible and applies to more general settings, e.g., gravity coupled to matter fields, as in charged Kerr-Newman-AdS BHs, which exhibit richer thermodynamic behaviors \cite{Yang:2025xck}. Furthermore, the implications of our framework for the thermodynamics of dual field theories in holographic setups also represent an interesting direction worthy of future investigation \cite{Chen:2024pyy}.

\section*{Acknowledgements}
YX is grateful to the Higgs Centre for Theoretical Physics at the University of Edinburgh for providing research facilities and hospitality during the visit. YX is supported in part by the National Natural Science Foundation of China (Grant No.12475048), the Hebei Natural Science Foundation (Grant No.A2024201012), the Science Research Project of Hebei Education Department (Grant No.JCZX2026019) and the China Scholarship Council (Grant No.202408130101). This work is also supported in part by the National Natural Science Foundation of China (Grants No.12575047, No.12447182, No.12475056, No.12375058, No.12361141825, No.12247101, and  No.12035016).





\section*{Appendix A: Holographic Renormalization Approach for the Thermodynamic Volume}

Two standard regularization schemes are used to extract physical results from divergent AdS quantities: redshifted background subtraction \cite{hawkingpage,Gubser:1998nz,Gibbons:2004ai} and holographic renormalization \cite{Balasubramanian:1999re,Bianchi:2001kw,deHaro:2000vlm}. We adopt background subtraction in the main text due to its simplicity of implementation. In contrast, holographic renormalization is an alternative scheme rooted in the holographic principle; its physical rationale lies in rendering the divergences of AdS quantities finite by introducing local boundary counterterms from the perspective of the boundary theory. These counterterms are often difficult to derive for higher-derivative actions. That said, holographic renormalization naturally aligns with AdS/CFT and offers unique advantages. We therefore outline the key steps for obtaining the geometric expression for $\mathcal{V}_i$ within this scheme.

In holographic renormalization, the divergences in $\int_{S_\infty} (\boldsymbol{Q} + \xi \cdot \boldsymbol{B})$ (where $\boldsymbol{Q}$ is the Noether charge and $\boldsymbol{B}$ the boundary term) are canceled by adding a counterterm $\xi \cdot \boldsymbol{S}_{ct}$. No reference background is needed, and divergences are removed purely via counterterms. Then, following the analysis in the main text, we identify the regularized mass and angular momentum as:
\begin{align}
   & M = \int_{S_\infty}^{(\mathrm{BH})} \left[ \boldsymbol{Q}_{\xi_t} + \Omega_\infty \boldsymbol{Q}_{\xi_\phi} + \xi_t \cdot (\boldsymbol{B} + \boldsymbol{S}_{ct}) \right], \tag{A1}  \\
   & J = -\int_{S_\infty}^{(\mathrm{BH})} \boldsymbol{Q}_{\xi_\phi},  \tag{A2}
\end{align}
and the thermodynamic volume as:
\begin{equation}
\begin{split}
       \mathcal{V}   & = -\int_{\Sigma}^{(\mathrm{BH})}\xi_t \cdot\boldsymbol{\epsilon}+ \int_{S_\infty}^{(\mathrm{BH})} \xi_t\cdot \big(\boldsymbol{F}[\frac{\partial \phi}{\partial \alpha_0}]+ \frac{\partial \boldsymbol{S}_{ct}}{\partial \alpha_0}\big)\\
       & =\int_{S_\infty}^{(\mathrm{BH})}\bigg[ -\boldsymbol{\psi}_{\xi_t}+ \xi_t\cdot (\boldsymbol{F}[\frac{\partial \phi}{\partial \alpha_0}]+  \frac{\partial \boldsymbol{S}_{ct}}{\partial \alpha_0})\bigg],
\end{split}\tag{A3}\label{a3}
\end{equation}
where $\alpha_0\equiv-\Lambda/(8\pi)$. The second line simply rewrites the formula in a more holographically intuitive form. Concretely, using the trick from Refs.~\cite{Kastor:2009wy,Cvetic:2010jb,Hajian:2023bhq}, i.e., introducing the Killing potentials $\boldsymbol{\psi}_{\xi}$ satisfying $d\boldsymbol{\psi}_{\xi} = \xi \cdot \boldsymbol{\epsilon}$, the volume integral in the first line is converted to a boundary integral. Gauge freedom further allows us to set the horizon contribution of $\boldsymbol{\psi}_\xi$ to zero, rendering Eq.~\eqref{a3} a pure boundary integral at infinity. However, for practical computations, the first line of Eq.~\eqref{a3} remains more convenient. 
The other quantities $\mathcal{V}_m$ conjugate to $\alpha_m$ can be obtained analogously.

\section*{Appendix B: Redshifted vs. Unredshifted Background Subtraction, with a Comparison to Existing Literature}

In the literature \cite{Kastor:2009wy,Dutta:2022wbh,Xiao:2023lap,Jacobson:2018ahi}, a commonly used formula is given by
\begin{equation}
V\equiv -\int_{\Sigma}^{(\mathrm{reg.})}\xi_t\cdot\boldsymbol{\epsilon}\equiv-\big(\int_{\Sigma}^{(\mathrm{BH})}\xi_t\cdot\boldsymbol{\epsilon}-\int_{\Sigma}^{(\mathrm{AdS_0})}\xi_t\cdot\boldsymbol{\epsilon}\big).\,\tag{B1}\label{b1}
\end{equation}
which is often recast as a boundary integral using Killing potentials $\boldsymbol{\psi}_{\xi}$ defined via $d\boldsymbol{\psi}_{\xi} = \xi \cdot \boldsymbol{\epsilon}$. This formula yields the naive volume of a BH, e.g., $V=\frac{4\pi}{3}r_h^3$ in $D=4$ dimensions. As noted in the Introduction, it satisfies $\mathcal{V} =V$ for a Schwarzschild–AdS BH, but $\mathcal{V} \neq V$ in generic cases.

We note that Eq.~\eqref{b1} relies on \emph{unredshifted} pure AdS spacetime (obtained from the BH metric by setting $M=0$) for background subtraction, whereas all results in our main text use a \emph{redshifted} pure AdS spacetime (setting $M=0$ and introducing a redshift factor to match the boundary conditions). For clarity, we use the notation ``$\int_{\Sigma}^{(\mathrm{reg.})}$'' and ``$\mathrm{AdS_0}$'' for the unredshifted scheme, in contrast to ``$\int_{\Sigma}^{(\mathrm{Reg.})}$'' and ``$\mathrm{AdS}$'' for the redshifted scheme.

We first emphasize that the validity of physical relations is unaffected by the choice of backgrounds. This can be illustrated by a simple identity: if $A+B=C$ and the background satisfies $A_0+B_0=0$, then $(A-\lambda A_0)+(B-\lambda B_0)=C$ holds for any $\lambda$ (with or without the redshift factor). Different values of $\lambda$ do not change $C$, but simply modify the physically meaningful decomposition of the quantities. In practice, both backgrounds are useful for specific purposes: starting from $\int_{\Sigma} (d\boldsymbol{Q}_\xi-\xi \cdot \boldsymbol{L})=0$, unredshifted background subtraction leads to the Smarr relation, while redshifted subtraction yields the thermodynamic relation $F=M-TS+\cdots$.

 Existing literature tends to express the thermodynamic volume as the naive volume $V$ in Eq.~\eqref{b1} plus a correction term $\Delta V$. This suggests decomposing $\mathcal{V}_i$ as $\mathcal{V}_i=V_i+\Delta V_i$, where $V_i\equiv -\int_{\Sigma}^{(\mathrm{reg.})}L_i\xi_t\cdot\boldsymbol{\epsilon}$ is the direct generalization of Eq.~\eqref{b1}. A critical drawback of such approaches is that they are forced to use unredshifted background subtraction for regularization, since the definition of $V$ is already tied to this scheme. We now explain that unredshifted background subtraction inherently fails to yield a geometric expression for $\Delta V_i$ (and hence for $\mathcal{V}_i$).

Consider a general higher-derivative gravity theory described by the Lagrangian $\boldsymbol{L}=\big(\frac{R}{16\pi}+\sum_i\alpha_i L_i\big)\boldsymbol{\epsilon}$ and a non-rotating BH solution. A standard strategy to derive the geometric expression of $\mathcal{V}_i$ is to compare thermodynamic relations with their geometric counterparts. One approach is to compare the ``thermodynamic'' Smarr relation $M=2TS+2\sum_i(i-1)\alpha_i \mathcal{V}_i$ with the ``geometric'' Komar integral relation \cite{Kastor:2009wy,Cvetic:2010jb}. This gives the correspondence
\begin{equation}
\begin{adjustbox}{width=\linewidth}
$\displaystyle
\overbrace{2\int_{S_\infty}^{(\mathrm{reg.})} \bs{Q}_{\xi_t}}^{M\!-\!2\sum_i(i-1)\alpha_i \Delta V_i}\!-\!\overbrace{2\int_{S_h}\bs{Q}_{\xi_t}}^{2TS}= \overbrace{-2\sum_i(i\!-\!1)\alpha_i \int_\Sigma^{(\mathrm{reg.})}L_i\,\xi_t \cdot\bs{\epsilon}}^{2\sum_i(i-1)\alpha_i V_i}.$
\end{adjustbox}
\tag{B2}\label{b2}
\end{equation}
An alternative approach is to compare the first law $\tilde{\delta} M=T\tilde{\delta} S+\mathcal{V}_i\tilde{\delta} \alpha_i$ with the Iyer--Wald charge identity regularized via unredshifted background subtraction \cite{Xiao:2023lap}, which yields
\begin{equation}
\begin{adjustbox}{width=\linewidth}
$\displaystyle
\overbrace{\int_{S_\infty}^{(\mathrm{reg.})} \left(\tilde{\delta} \boldsymbol{Q}_{\xi_t}   -  \xi_t  \cdot  \boldsymbol{\Theta}[\tilde{\delta} \phi] \right)}^{\tilde{\delta} M-\sum_i\Delta V_i\tilde{\delta}\alpha_i} -\overbrace{\int_{S_h}^{(\mathrm{BH})} \tilde{\delta} \boldsymbol{Q}_{\xi_t}}^{T\tilde{\delta} S}=\overbrace{\sum_i \tilde{\delta}\alpha_i \int^{(\mathrm{reg.})}_\Sigma (-L_i)\xi\cdot \boldsymbol{\epsilon}}^{\sum_i V_i\tilde{\delta}\alpha_i}.$
\end{adjustbox} 
\tag{B3}\label{b3}
\end{equation}
In both approaches, all thermodynamic contributions associated with $\Delta V_i$ are conflated into a single geometric term in Eqs.~\eqref{b2} and \eqref{b3}. As a result, one cannot proceed further with the ``geometric-thermodynamic quantity correspondence'' to isolate a separate geometric expression for each $\Delta V_i$. Moreover, introducing Killing potentials to rewrite $\int L_i \xi_t\cdot \boldsymbol{\epsilon}$ in Eqs.~\eqref{b2} and \eqref{b3} as boundary integrals \cite{Kastor:2009wy,Cvetic:2010jb,Hajian:2023bhq} provides no improvement; it merely converts the problem of term conflation into one of gauge ambiguity.

In our main text, adopting the standard \emph{redshifted} scheme naturally leads to a clean separation of all relevant terms, thus avoiding the problem of term mixing. In a sense, our work implies that the physically natural decomposition is $\mathcal{V}_i=\mathcal{V}_i^{(1)}+\mathcal{V}_i^{(2)}$, rather than $\mathcal{V}_i=V_i+\Delta V_i$. The latter decomposition is artificially motivated by the appearance of $V$ in the simplest Schwarzschild–AdS case, but in general there is no sound physical basis to split $\mathcal{V} $ into a naive volume term $V$ and an ad hoc correction.

\vfill


\vfill

\end{document}